\documentclass[conference]{IEEEtran}
\pdfoutput=1
\makeatletter
\def\ps@headings{%
\def\@oddhead{\mbox{}\scriptsize\rightmark \hfil \thepage}%
\def\@evenhead{\scriptsize\thepage \hfil \leftmark\mbox{}}%
\def\@oddfoot{}%
\def\@evenfoot{}}
\makeatother
\usepackage{subfigure}
\usepackage{cite}
\usepackage{empheq}
\ifCLASSINFOpdf
\else
\fi

\usepackage{amsmath}
\usepackage{verbatim}

\usepackage{algorithmic}
\usepackage{algorithm}
\usepackage{bm}
\usepackage{amsfonts,amssymb}
\usepackage{graphicx} 
\usepackage{float} 
\usepackage{subfig} 
\begin{document}
\captionsetup{font={small},singlelinecheck=off}
%
\title{\huge Wireless Indoor  Simultaneous Localization and Mapping Using Reconfigurable Intelligent Surface}
%
%
%
\author{\IEEEauthorblockN{Ziang~Yang\IEEEauthorrefmark{1},
Haobo~Zhang\IEEEauthorrefmark{2},
Boya~Di\IEEEauthorrefmark{2},
Hongliang~Zhang\IEEEauthorrefmark{3},
Kaigui~Bian\IEEEauthorrefmark{4},
and~Lingyang~Song\IEEEauthorrefmark{2}}
\IEEEauthorblockA{\IEEEauthorrefmark{1}Department of Machine Intelligence, Peking University, Beijing, China.}
\IEEEauthorblockA{\IEEEauthorrefmark{2}Department of Electronics, Peking University, Beijing, China.}
\IEEEauthorblockA{\IEEEauthorrefmark{3}Department of Electrical and Computer Engineering, Princeton University, NJ, USA.}
\IEEEauthorblockA{\IEEEauthorrefmark{4}Department of Computer Science, Peking University, Beijing, China.}
}
\maketitle
\begin{abstract}
Indoor wireless simultaneous localization and mapping (SLAM) is considered as a promising technique to provide positioning services in future 6G systems. However, the accuracy of traditional wireless SLAM system heavily relies on the quality of propagation paths, which is limited by the uncontrollable wireless environment. In this paper, we propose a novel SLAM system assisted by a reconfigurable intelligent surface (RIS) to address this issue. By configuring the phase shifts of the RIS, the strength of received signals can be enhanced to resist the disturbance of noise. However, the selection of phase shifts heavily influences the localization and mapping phase, which makes the design very challenging. To tackle this challenge, we formulate the RIS-assisted indoor SLAM optimization problem and design an error minimization algorithm for it. Simulations show that the RIS assisted SLAM system can decrease the positioning error by at least 31\% compared with benchmark schemes. 
\end{abstract}


%
\IEEEpeerreviewmaketitle
\section{Introduction}
To enable location and sensing based services in future 6G networks such as \cite{8624565} and \cite{8926369}, simultaneous localization and mapping~(SLAM) technique is a promising solution. Specifically, the SLAM technique uses sensors such as camera, laser, or antennas to sense the surrounding area and estimates the location of the agent equipped with these sensors. Recently, wireless SLAM which uses antennas to exploit the multipath effect of the wireless signals for SLAM has attracted much attention\cite{7487067}. Unlike vision SLAM which uses cameras and fails in dark environment, wireless SLAM is robust to light conditions. Besides, the wireless SLAM is cost effective because the antennas are cheap and existing wireless infrastructures can be leveraged.

In the literature, various wireless SLAM methods have been discussed. For example, in \cite{7487067}, the authors proposed Channel-SLAM, which considers both scattering and reflection effects of multipath components (MPCs). A range-only SLAM method using time of arrival (TOA) was designed in \cite{8823946}. However, the performance of aforementioned SLAM systems highly depends on the quality of MPCs. If the MPC amplitude is not strong enough compared to the noise, the estimated TOA and angle of arrival (AOA) of this MPC will have significant errors, leading to a lower SLAM accuracy\cite{6232446}.

Fortunately, the reconfigurable intelligent surface (RIS) has been proposed as a promising tool to control the wireless environment, which can be used to address this issue\cite{9110889},~\cite{9193909}. An RIS is a type of plane material consisting of many homogeneous elements, which can be coated on the surface of walls or ceilings. The RIS has the capability to shape radio waves impinging upon the surface by changing the phase shift of each element. In this way, the indoor propagation environment can be customized according to the requirements of SLAM systems, and thus, the accuracy can be improved.

In this paper, we consider a SLAM system assisted by the RIS. While the agent moves in the room, it transmits signals and receives signals carrying information of the surrounding environment at the same time, and the RIS enhances the strengths of received signals to reduce the impact of noise. New challenges have arisen in the RIS-assisted SLAM system. The first challenge lies in the way of collaboration between the RIS and the mobile agent. Secondly, due to the heavy impact of the selection of phase shifts on the localization and mapping algorithm, it is challenging to select phase shift of the RIS that can significantly improve the performance of wireless SLAM system from enormous number of possible phase shifts. 

In response to the above challenges, we propose an RIS-assisted SLAM protocol to coordinate RIS and the mobile agent, and formulate an optimization problem aiming at minimizing the agent positioning error. The optimization problem is solved by an error minimization algorithm, which consists of two sequential phases: phase shift optimization phase and localization and mapping estimation (LME) function optimization phase. The phase shift is optimized by an iterative optimization approach, and LME function is optimized by adjusting the weight of MPCs according to the amplitudes of the received signals optimized in the previous phase. Simulation results show that the positioning error can be reduced by at least 31\% using the proposed scheme.

The rest of the paper is organized as follows. In Sec. \ref{sec:model}, we provide the indoor SLAM scenario and the received signal model. An RIS-aided SLAM protocol is proposed in Sec. \ref{sec:protocol}. In Sec. \ref{sec:algorithm}, an optimization problem aiming at improving the SLAM performance is proposed, and an error minimization algorithm to efficiently solve the formulated problem is designed. In Sec. \ref{sec:sim}, we present the simulation results and discussions. Finally, conclusions are drawn in Sec. \ref{sec:conclusion}.

\section{System Model}
\label{sec:model}
In this section, we first introduce the indoor SLAM scenario, and then provide the received signal model.
\subsection{Indoor SLAM Scenario}
\begin{figure}[t]
    \centering
    \includegraphics[width=0.33\textwidth]{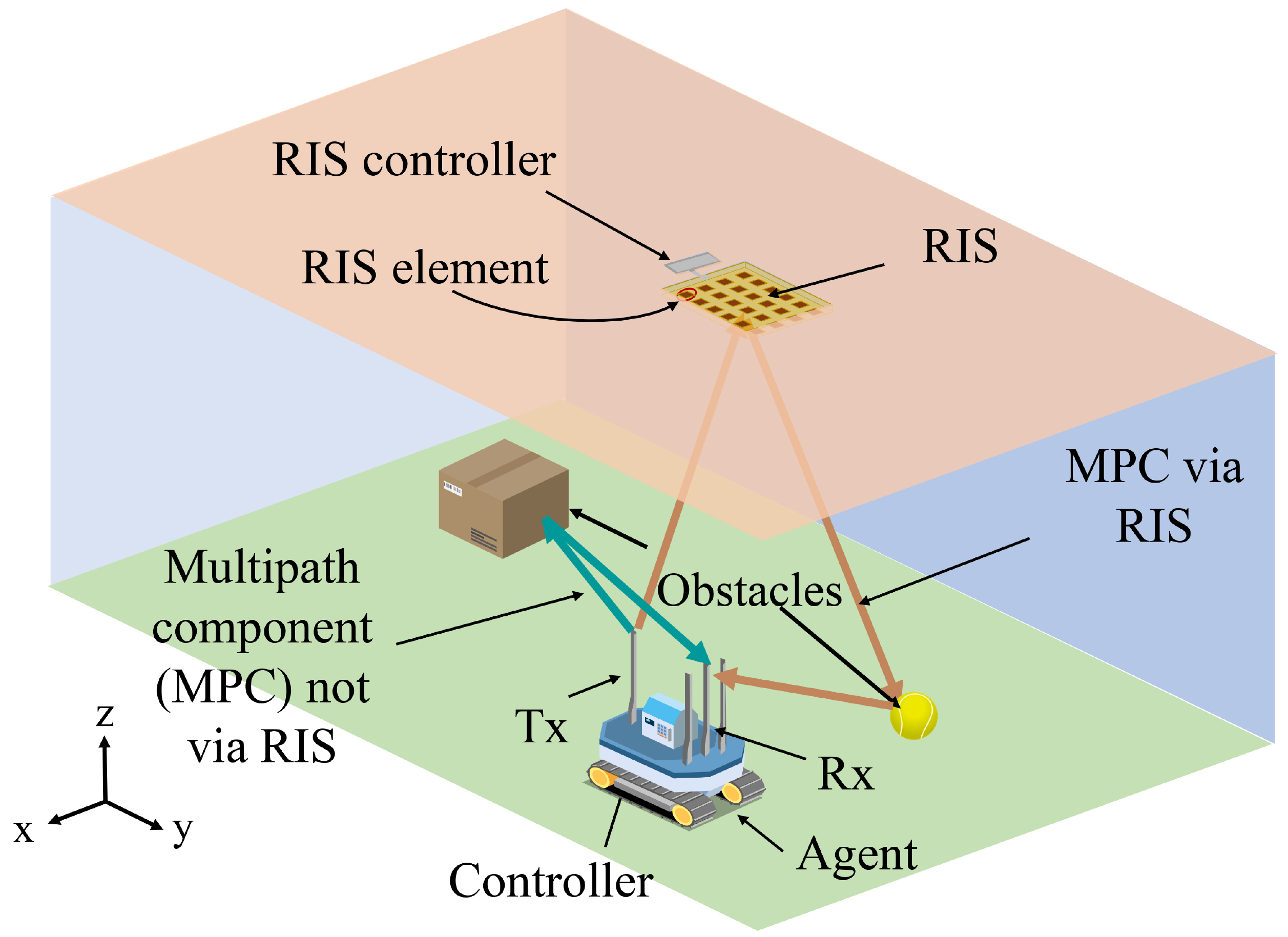}
    \caption{System model for the RIS-assisted wireless SLAM.}
    \label{fig:scenario}
\end{figure}

 As shown in Fig. \ref{fig:scenario}, we consider an indoor SLAM scenario consisting of an RIS on the ceiling, an RIS controller, several obstacles, and a mobile agent equipped with a controller, a single-antenna transmitter (Tx) and a multiple-antenna receiver (Rx). As the agent moves in the room, it will transmit signals and analyze the received signals in order to locate itself and map the surrounding environment at the same time.
 
Specifically, to obtain its location and the map information, the agent first communicates with the RIS controller to adjust the RIS phase shifts. Next, the agent simultaneously emits signals to the environment and records the received signals. The received signals contain MPCs produced by the reflection and scattering of the obstacles and the RIS. The location of the mobile agent and environment information can be extracted from these MPCs. In addition, by leveraging the ability of the RIS to adjust wireless channels, the power of MPCs via the RIS can be strengthened, bringing better SLAM performance.

 \subsection{Received Signal Model}
 \label{sec:obs model}
Assume that the Rx is equipped with $L$ antennas, and the signal $s(t)$ emitted by the Tx reaches the Rx via $P(t)$ multipath channels. Thus, the signals received by the $l$-th Rx antenna at time $t$ can be written as\cite{goldsmith2005wireless}
\begin{equation}
\begin{small}
\label{eq:all}
    y_l(t)=\int_{-\infty}^\infty\sum_{i=0}^{P(t)-1}h_{i,l}(t,\tau)s(t-\tau)d\tau+n(t),
 \end{small}
\end{equation}where $h_{i,l}(t,\tau)$ denotes the channel impulse response (CIR) of the $i$-th multipath channel, and $n(t)$ denotes white circular symmetric normal distributed receiver noise with variance $\sigma_n^2$.

In the following, we model the impulse response of channels via the obstacles and RIS. Two types of obstacles are considered, i.e. reflectors and scatters \cite{7487067}. Let $\mathcal{R}$ and $\mathcal{S}$ denote the sets of all the reflectors and scatterers, respectively. The reflectors are smooth surfaces like walls where the incident signals change the propagation direction following the reflection law. As illustrated in Fig.2~(a), channel~1 is the signal path from the Tx to the Rx via reflector $R_1$, and the incident angle $\theta_i$ is equal to the reflection angle $\theta_r$. The CIR of channel~1 from Tx antenna to the $l$-th Rx antenna can be expressed as\cite{goldsmith2005wireless}:
\begin{equation}
\label{a1}
    h_{1,l}(t,\tau)=\frac{\lambda R\sqrt{G}e^{-j2\pi d_1/\lambda}}{4\pi d_1}b_l(\phi_1(t))\delta(\tau-d_1/c),
\end{equation}
where $G$ is the antenna gain, $\lambda$ is the wavelength of the signal, $\phi_1(t)$ denotes the AOA, $b_l(\phi_1(t))$ is the response of the $l$-th receiving antenna with regard to the phase center,  $R$ is the reflection coefficient of $R_1$, $d_1$ is the propagation length of channel~1, and $\delta(\tau)$ denotes the delta function. Note that this CIR is equivalent to the CIR from the virtual transmitter~(VT), which is the mirror image of the TX.
\begin{figure} \centering    
\subfigure[] {
\includegraphics[width=0.43\columnwidth]{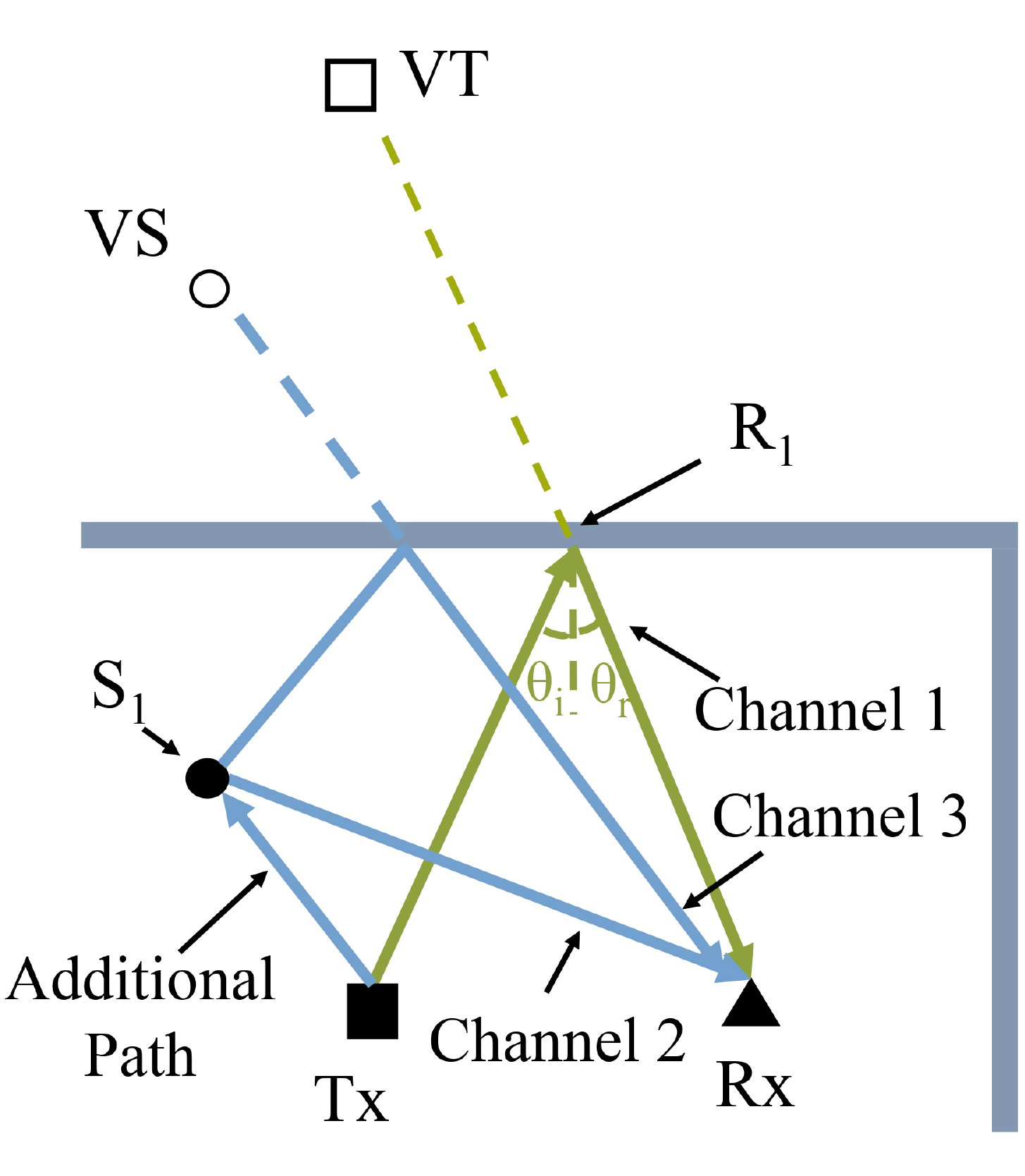}  
}     
\subfigure[] { 
\includegraphics[width=0.42\columnwidth]{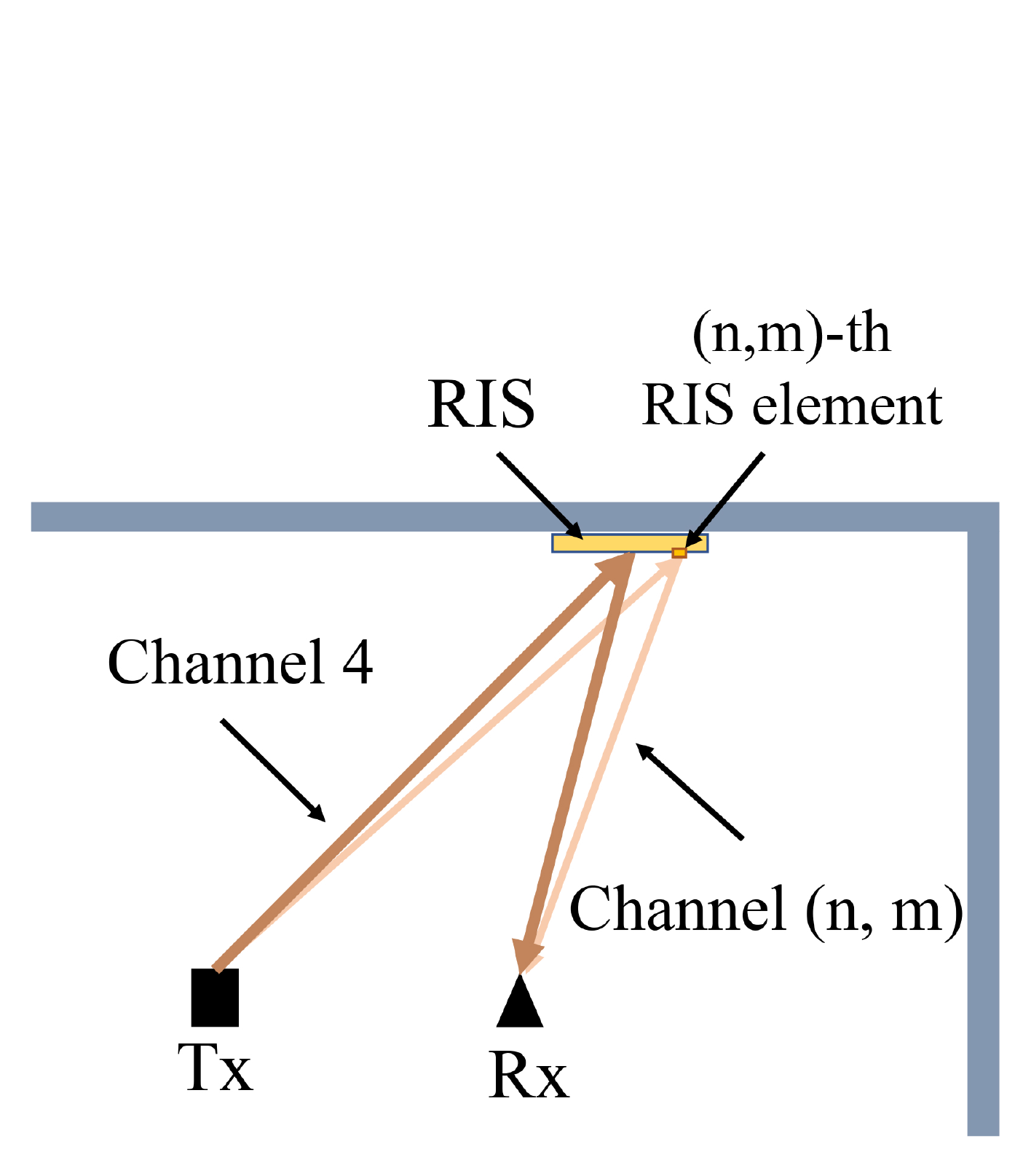}     
}    
\caption{Illustration of the multipath channels.}     
\end{figure}

Different from the reflectors, the scatters are small objects that will radiate the energy of incident signals in all directions. As shown in Fig. 2(a), channel~2 is the signal path from the Tx to the Rx via scatterer $S_1$, and its CIR from Tx antenna to the $l$-th Rx antenna can be expressed as is \cite{goldsmith2005wireless}
\begin{equation}
\label{a2}
    h_{2,l}(t,\tau)=\frac{\lambda\sqrt{G\sigma_1}e^{-j2\pi d_2 /\lambda}}{(4\pi)^{3/2}d_{T,S_1}d_{S_1,R}}b_l(\phi_2(t))\delta(\tau-d_2/c),
\end{equation}
where $\sigma_1$ is the radar cross-section of the scatterer $S_1$, $d_{T,S_1}$ is the distance between Tx and scatterer $S_1$, $d_{S_1,R}$ is the distance between $S_1$ and Rx, and $d_2 = d_{T, S_1} + d_{S_1, R}$ is the propagation length of channel~2.

Besides, channel~3 is the propagation path from Tx to the Rx via scatterer $S_1$ and reflector $R_1$, and its CIR can be calculated based on (\ref{a1}) and (\ref{a2}). Similar to the MPC via channel~1, the MPC via channel~3 can be treated as being transmitted by a virtual scatterer~(VS) with an additional delay $d_{T,S_1}$ introduced by the additional path\cite{7487067}. The VS is also the mirror image of physical scatterer(PS) $S_1$ with the additional path same to channel~2. Similar to the definition of PS and VS, we can define RIS and VRIS. Since the locations of PS, VS, RIS and VRIS are fixed, they can be viewed as \emph{landmarks} for localization and mapping, which will be introduced in Sec. \ref{sec:algorithm}.

As for RIS, suppose it is composed of $N\times M$ elements. channel~$\rm {(n,m)}$ in Fig. 2(b) is the multipath channel~from the Tx to the $l$-th Rx antenna via the element in row $n$ and column $m$, and its gain can be written as\cite{9206044}:
\small
\begin{align}
\label{eq:single}
    \alpha_{n,m,l}&=\frac{\lambda\sqrt{GG_{R}F(\theta_{n,m}^t,\phi_{n,m}^t)F(\theta_{n,m,l}^r,\phi_{n,m,l}^r)d_xd_y}}{8\pi^{3/2}r_{n,m}^tr_{n,m,l}^r}\notag\\
    &\times e^{\frac{-j2\pi(r_{n,m}^t+r_{n,m,l}^r)}{\lambda}} Ae^{-j\xi_{n,m}},
\end{align}\normalsize
where $G_{R}$ denotes the RIS element gain, $F(\theta,\phi)$ denotes the radiation pattern of RIS in direction $(\theta, \phi)$. $(\theta_{n,m}^t,\phi_{n,m}^t)$ and $(\theta_{n,m,l}^r,\phi_{n,m,l}^r)$ denote the directions from the RIS to the Tx and the Rx, respectively. $d_x$ and $d_y$ are width and length of an RIS element, respectively. $A$ is the reflection coefficient of the RIS element, and $\xi_{n,m}$ is the phase shift caused by the reflection of the (n, m)-th element. We assume that each element has $H$ possible phase shifts with uniform interval $\Delta\theta=2\pi/H$. Thus, we have $\xi_{n,m}=h_{n,m}\Delta\theta$, where $h_{n,m}\in\{1,2,...,H\}$. $r_{n,m}^t$ and $r_{n,m,l}^r$ denote the distances from the $(n,m)$-th element to the Tx and the $l$-th antenna of Rx, respectively.

As the MPCs from the Tx to the Rx via different RIS elements will overlap at the Rx\footnote{The maximum delay of MPCs from the Tx to the Rx via different RIS elements can be approximated by $D/c$,  where $D$ is the size of the RIS. In this paper, we use an RIS with $6 \times 6$ elements and working frequency at $10GHz$, and the maximum delay is about $0.3$ns. Since the accuracy of mainstream devices is $0.5$ns\cite{7102674}, the MPCs via different elements are indistinguishable at the Rx.}, the $N\times M$ channels from Tx to the Rx via $N \times M$ elements can be substituted by channel~4 in Fig. 2. By applying the far-field model\footnote{The far-field region of an RIS is $d  > 2D^2/\lambda$\cite{9206044}, where $d$ is the distance between the Tx/Rx and the RIS center. For the RIS used in this paper, its farfield region is $d > 0.54m$. Since the RIS is on the ceiling and the agent is moving on the ground, the Tx and Rx are always in the farfield of the RIS.}, we have proved in Appendix A that the CIR of channel~4 is given by
\small
\begin{equation}
\label{a3}
\begin{split}
    h_{4,l}(t,\tau)&\approx\frac{\lambda\sqrt{GG_{RIS}F(\theta_t,\phi_t)F(\theta_r,\phi_r)d_xd_y}}{(4\pi)^{3/2}d_{T,RIS}d_{RIS,R}}\\
    &\times\sum_{n=1}^{N}\sum_{m=1}^{M}e^{\frac{-j2\pi(d_3- \bm{z}_{n,m}\cdot\bm{z}_t-\bm{z}_{n,m}\cdot\bm{z}_r)}{\lambda}}Ae^{-j\xi_{n,m}}\\
    &\times b_l(\phi_4(t))\delta(\tau-d_4/c) ,
    \end{split}
\end{equation}\normalsize
where $(\theta_t,\phi_t)$ and $(\theta_r,\phi_r)$ denote the directions from the RIS center to the Tx and Rx, respectively. $d_{T, RIS}$ and $d_{RIS, R}$ are the distances from RIS center to Tx and Rx, respectively. $d_3 = d_{T, RIS} + d_{RIS, R}$ is the propagation length of channel~3. $\bm{z}_{n,m}$ denotes the position of the $(n,m)$-th element. $\bm{z}_t$ and $\bm{z}_r$ denote unit vectors in directions $(\theta_t,\phi_t)$ and $(\theta_r,\phi_r)$, respectively.

\section{RIS Assisted SLAM Protocol}
\label{sec:protocol}
 \begin{figure}[t]
    \centering
    \includegraphics[width=0.42\textwidth]{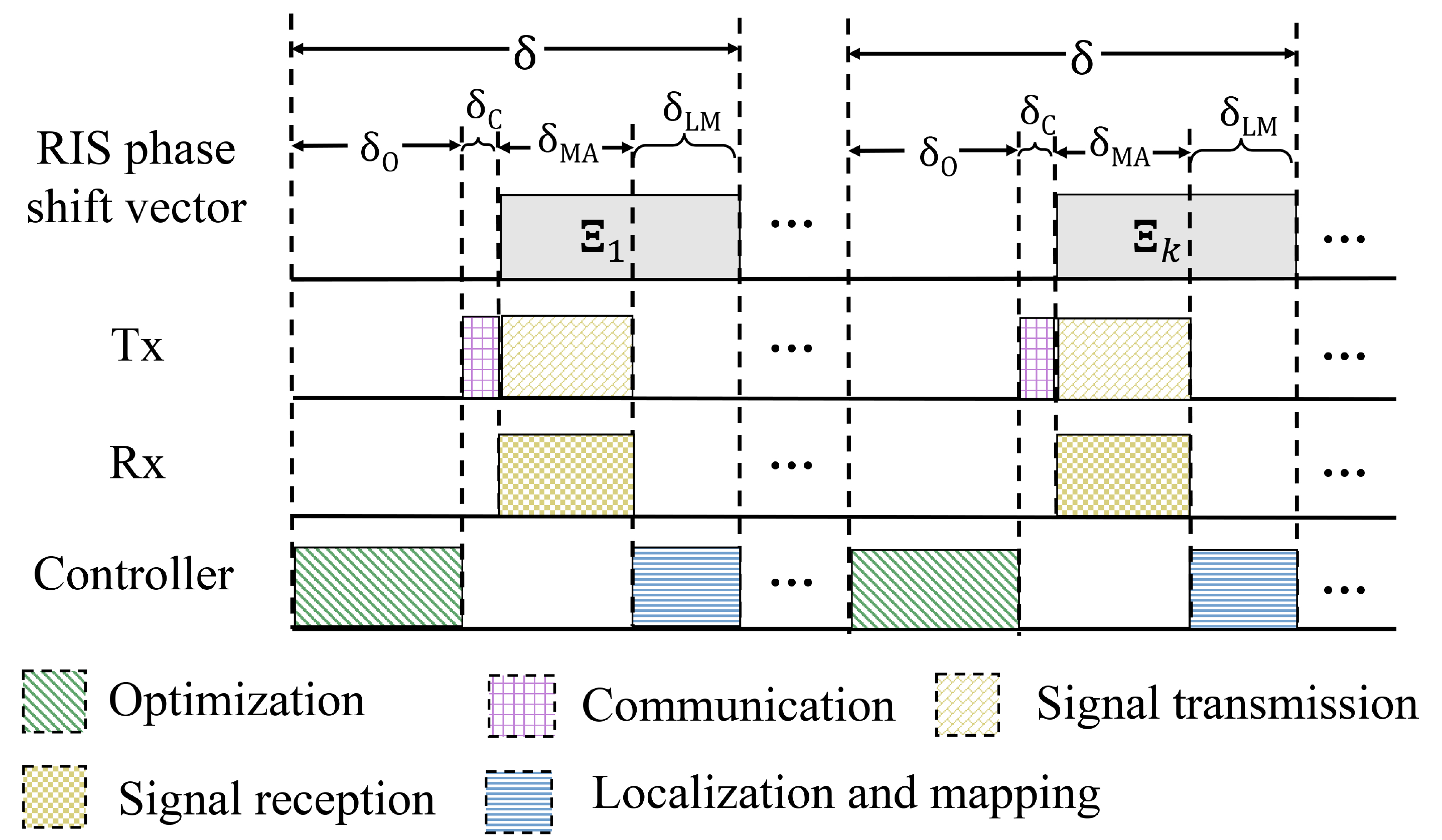}
    \caption{RIS assisted SLAM protocol.}
    \label{fig:protocol}
\end{figure}
In this section, we propose an RIS assisted SLAM protocol, where the RIS and the agent are coordinated to improve the SLAM performance. The process of the positioning protocol is illustrated in Fig. \ref{fig:protocol}.
We divide the timeline into cycles with duration $\delta$. In each cycle, the optimization, communication, measurement acquisition, localization and mapping steps are conducted sequentially.

\textbf{Optimization}: In the first $\delta_O$ seconds of the $k$-th cycle, the agent selects the optimal phase shifts $\boldsymbol{\Xi}^k=\{\xi_{1,1}^k,\xi_{1,2}^k,...,\xi_{N,M}^k\}$ and the LME function $f^k$ for this cycle based on collected information in previous cycles. The function $f^k$ can estimate the agent position $\bm{p}_a^k$, the agent velocity $\bm{v}^k$, and the map $\mathcal{M}^k$ given received signal $y^k(t)$, phase shifts $\boldsymbol{\Xi}^k$, the agent position $\bm{p}_a^{k-1}$, the velocity $\bm{v}_a^{k-1}$, and map $\mathcal{M}^{k-1}$. Here, the map $\mathcal{M}^k$ denotes the estimated locations of reflectors, scatters, and the RIS in the $k$-th cycle. 

\textbf{Communication}: In the next $\delta_C$ seconds, the agent transmits a signal carrying the information of $\boldsymbol{\Xi}^k$ to the RIS controller, and the phase shifts of RIS will be set as $\boldsymbol{\Xi}^k$.

\textbf{Measurement Acquisition}: After the RIS phase shifts are adjusted, the agent transmits signal $s(t)$ for SLAM, and the Rx records the received signal $y^k(t)$ at the same time.

\textbf{Localization and Mapping}: In this step, the location of the agent $\bm{p}_a^k$, the velocity $\bm{v}_a^k$, and the map $\mathcal{M}^k$ are estimated using received signal $y^k(t)$, map $\mathcal{M}^{k-1}$, the position of the agent $\bm{p}_a^{k-1}$, the velocity $\bm{v}_a^{k-1}$, and the optimal $f^k$. 
\section{Localization Error Minimization}
\label{sec:algorithm}
In this section, we first formulate the optimization problem aiming at minimizing the agent position error, then propose an error minimization algorithm to solve the formulated problem.
\subsection{Problem Formulation}
The objective of the optimization problem in each cycle is to minimize the agent localization error. Let $\bm{p}_a^k$ and $\hat{\bm{p}}^k_a$ denote the real and estimated agent location in the $k$-th cycle, respectively, the optimization problem can be formulated as
\begin{subequations}
\begin{align}
\label{allopt}
    &\min_{\boldsymbol{\Xi}^k, f^k}||\bm{p}_a^k-\hat{\bm{p}}_a^k||^2, \\   &s.t.~~h^k_{i,j} \in \{1,...,H\}, i=1,2,...,N,j=1,2,...,M.
\end{align}
\end{subequations}

Notice that $\boldsymbol{\Xi}^k$ and $f^k$ are heavily coupled, so it is hard to jointly optimize them. As a result, we decompose (\ref{allopt}) into two sub-problems by separating the  optimization of $\boldsymbol{\Xi}^k$ and $f^k$. These two sub-problems are described as follows:

 \textbf{Phase shift optimization}: Given $f^k$, the optimization sub-problem for $\boldsymbol{\Xi}^k$ can be given by
\begin{subequations}
\begin{align}
\label{alloptimization}
    &\min_{\boldsymbol{\Xi}^k}||\bm{p}_a^k-\hat{\bm{p}}_a^k||^2, \\   &s.t.~~h^k_{i,j} \in \{1,...,H\}, i=1,2,...,N,j=1,2,...,M.
\end{align}
\end{subequations}

\textbf{LME function optimization}: Given a phase shift vector, the optimization sub-problem of $f^k$ can be expressed by 
\begin{equation}
    \min_{f^k}||\bm{p}_a^k-\hat{\bm{p}}_a^k||^2.
\end{equation}
Since the real location of the agent is unknown, it is difficult to directly optimize RIS phase shifts with the positioning error (7a) as the optimization objective. Alternatively, for the optimization of RIS phase shifts, we use Cramer-Rao lower bound (CRLB) to approximate the positioning error, which is widely used in assessing the performance of the a SLAM system\cite{6363827}. Suppose that the variable to be estimated is denoted by $\boldsymbol{\iota}$, and $\hat{\boldsymbol{\iota}}$ denotes the estimation of $\boldsymbol{\iota}$, the CRLB can be expressed as ${\rm CRLB}(\hat{\boldsymbol{\iota}})= \bm{J}^{-1}(\boldsymbol{\iota})$, where $\bm{J}(\cdot)$ is the Fisher information matrix (FIM)\cite{6363827}. The optimization of $\boldsymbol{\Xi}^k$ is discussed in detail in Sec.~\ref{pso}. For the optimization of $f^k$, we adjust the parameters of it according to the amplitude of the received signals to minimize the agent positioning error. Detailed optimization method of $f^k$ can be found in Sec.~\ref{lmeo}.
\subsection{Algorithm Design}
An error minimization algorithm is proposed to solve the formulated problem in the $k$-th cycle. The algorithm contains two phases, i.e., the phase shift and LME function optimization phases, where $\boldsymbol{\Xi}^k$ and $f^k$ are optimized sequentially.
\subsubsection{Phase Shift Optimization Phase}
\label{pso}
As mentioned above, the objective of phase shift optimization is turned into the minimization of CRLB, so the phase shifts optimization problem in the $k$-th cycle can be expressed as:
\begin{subequations}
    \begin{align}
    \label{eq:target}
    &\min_{\boldsymbol{\Xi}^k} {\rm tr}(\bm{J}^{-1}(\boldsymbol{\eta}^k(\boldsymbol{\Xi}^k))_{[1:2] \times [1:2]}),\\
    &s.t.~~h^k_{i,j} \in \{1,...,H\}, i=1,2,...,N,j=1,2,...,M
    \end{align}
\end{subequations}
where $\rm tr(\cdot)$ denotes the trace of a matrix, $\boldsymbol{\eta}^k(\boldsymbol{\Xi}^k)=[(\bm{p}_{a}^k)^T, ({\rm real}(\bm{\alpha}^k))^T, ({\rm imag}(\bm{\alpha^k}))^T]^T$, where ${\rm real}(\bm{\alpha}^k)$ and ${\rm imag}(\bm{\alpha}^k)$ denote the real part and the imaginary part of the vector of channel gain $\bm{\alpha}^k$, respectively. $(\cdot)_{[1:2] \times [1:2]}$ denotes the submatrix which consists of the first two rows and columns of the original matrix. Since the position of the agent $\bm{p}_a^k = (r^k_x, r^k_y)$ is two-dimensional, where $r^k_x$ and $r^k_y$ denote the $x$ and $y$ coordinates of the position of the agent, respectively, the objective function (\ref{eq:target}) represents the sum of CRLB of variables $r^k_x$ and $r^k_y$. The expression of the objective function (\ref{eq:target}) is given by the following proposition:

\textbf{Proposition 1:} The objective (\ref{eq:target}) can be expressed as:
\begin{scriptsize}
\begin{equation}
\label{derived}
\begin{split}
    \bm{J}^{-1}(\boldsymbol{\eta}^k(\boldsymbol{\Xi}^k))_{[1:2] \times [1:2]}=\left(\frac{8\pi^2\zeta^2}{c^2}\!\sum_{i=0}^{P-1}\!\frac{|\alpha_i^k|^2}{N_0}\!\cdot\!\left[\begin{matrix}(\nu_i^k)^2& \nu_i^k\kappa_i^k\\ \nu_i^k\kappa_i^k&(\kappa_i^k)^2\end{matrix}\right]\right)^{-1},
    \end{split}
\end{equation}
\end{scriptsize}where $\zeta^2$ is the effective bandwidth of $s(t)$\cite{6363827}. $\nu^k_i$ and $\kappa_i^k$ are parameters related to the $i$-th channel. Specifically, assume that there are $N_i$ obstacles (or RIS) in the propagation channel, which divide the $i$-th channel into $N_i + 1$ segments, \small$\nu^k_i=\sum\limits_{j=1}^{N_i+1}x_j/\sqrt{x_j^2+y_j^2+z_j^2}$\normalsize, and  \small$\kappa^k_i=\sum\limits_{j=1}^{N_i+1}y_j/\sqrt{x_j^2+y_j^2+z_j^2}$\normalsize, where $(x_j,y_j,z_j)$ denotes the direction vector of the $j$-th segment.

\emph{Proof:} See Appendix B.

As the CRLB is non-convex, we design an optimization algorithm based on genetic algorithm (GA) to minimize CRLB, which is referred to as CRLB-GA. At the beginning, we randomly choose $K$ phase shift vectors to form the population set $\mathcal{P}$, which is denoted by $\mathcal{P}=\{\boldsymbol{\Xi}_1,...\boldsymbol{\Xi}_K\}$. A phase shift vector $\boldsymbol{\Xi}=[\xi_{1},...\xi_{N\times M}]$ in $\mathcal{P}$ is referred to as an individual, and an element $\xi_{i}$ in  $\boldsymbol{\Xi}$ is called a gene. Next, we iteratively update the individuals in the population. Specifically, the algorithm contains $C^A$ iterations, and the following three steps are executed sequentially in each iteration.

\textbf{Selection}: We first calculate the adaptability of each individual in set $\mathcal{P}$, and the adaptability of the $j$-th individual can be expressed as
\begin{equation}
\Gamma(\boldsymbol{\Xi}_j)=1/J^{-1}(\boldsymbol{\eta}(\boldsymbol{\Xi}_j))_{[1:2]\times[1:2]}
\end{equation}
Next, $Q$ individuals with the highest adaptability will be selected as elite individuals, denoted by $\mathcal{P}^e$, and they will be preserved for the next cycle.

\textbf{Gene cross}: In this step, we first choose two individuals as parents from $\mathcal{P}$, and the probability of choosing the $j$-th individual is is proportional to its adaptability. Then, we randomly choose a number $i \in [1, MN]$, and the genes from $i$ to $MN$ in the chosen parents are exchanged to generate two new individuals, which are referred to as the child individuals. This step will be repeated for $(K-Q)/2$ times, and the generated $K-Q$ child individuals are denoted by $\mathcal{C}$.

\textbf{Gene mutation}: For each child in set $\mathcal{C}$, its genes will mutate at a low probability. Specifically, each gene of a child may change to $h\Delta\theta, h \in \{1,...H\}$ at probability $p_m$. After that, the set $P_e$ and $\mathcal{C}$ will merge to generate the new population set $\mathcal{P}$.

\subsubsection{LME Function Optimization Phase}
\label{lmeo}
The localization and mapping function $f^k$ consists of two steps, i.e. the data association step and the position and map estimation step. In the first step, we link MPCs received by the Rx with PT, VT, PS, RIS, VS, or VRIS. In the next step, we design a particle filter based method to update the location of agent and the map, and the particle filter based method is optimized based on $\boldsymbol{\Xi}^k$ to promote its performance.

\textbf{Data Association}: 
First, we use a three-layer full-connected neural network to eliminate MPCs related to PT and VTs, this is because both PT and VTs are moving during the SLAM process, which cannot be utilized to locate the agent\cite{10.5555/1121596}.

Next, the MPCs are divided into $L$ groups based on their AOAs, and each cluster is corresponding to a PS, VS, RIS, or VRIS. The VS (or VRIS) can be associated to PS (or RIS) using the geometry method in\cite{8528841}. Here, the RIS (or VRIS) is treated as an ordinary PS (or VS). Besides, each group will be associated to a landmark based on the TOA and AOA~\cite{8528841}. 

Finally, we decide which \emph{landmark} is the RIS, which is necessary for the phase shift vector optimization phase. Let $p^k_{i, RIS}$ denote the probability that the $i$-th \emph{landmark} is RIS in the $k$-th cycle. Based on Bayes updating, we have
\small
\begin{equation}
    p_{i,RIS}^{k}=\frac{p_{i,RIS}^{k-1}p(\bm{\alpha}^k|L_i~\rm{is~RIS})}{\sum_{i=1}^Lp_{i,RIS}^{k-1}p(\bm{\alpha}^k|L_i~\rm{is~RIS})},
\end{equation}\normalsize
where $\bm{\alpha}^k$ denotes the amplitudes of MPCs extracted from the received signal $y^k$, $p(\bm{\alpha}^k|L_i~\rm{is~RIS})$ denotes the probability to receive $\bm{\alpha}^k$ if the $i$-th \emph{landmark} is RIS. When $k = 1$, we set $p^1_{i, RIS} = 1/L, \forall i$. The \emph{landmark} whose $p_{i,RIS}^k$ is the biggest will be regarded as the RIS in the $k$-th cycle.


\textbf{Position and Map Estimation}: 
The position and map estimation algorithm is based on the particle filter method \cite{7487067}. The basic idea of particle filter is to use a set of weighted particles to represent the probability distribution of the locations of the agent and \emph{landmarks}, and update the weights of particles based on the received signals. To update the weights of particles, we need to first extract TOAs $\bm{\tau}$ and AOAs $\bm{\phi}$ from MPCs obtained during measurement acquisition. Since MPCs with different amplitudes $\alpha$ have different estimation errors of TOAs $\tau$ and AOA $\phi$\cite{6232446}, we will optimize the weights of these MPCs in the particle filter to reduce the positioning errors and improve the algorithm performance. 

The particle filter consists of $N_{ap}$ agent particles. Each agent particle represents a possible pose of the agent and has multiple landmark particle sets. The $l$-th landmark particle set of the $i$-th agent particle contains $N_{lp,i,l}$ particles. Each particle represents a possible position of the \emph{landmark}. In the following, we show how the position and map estimation algorithm update the weights of the particles:

$\bullet$ Agent particle transition: In this step, the pose $\bm{r}_a^k=[\bm{p}_a^k, \bm{v}_a^k]$ is updated using the discrete time white noise acceleration model proposed in \cite{7487067}.

$\bullet$ MPC weight calculation: Assuming that $N_{MPC}$ MPCs are received in the temporal cycle, and the average amplitude for these MPCs is denoted by $\alpha_{ave}$, the weight $\rho_i$ for the $i$-th MPC can be calculated by $\rho_i=\alpha_i/\alpha_{ave}$.

$\bullet$ Weight update of landmark particles: Assuming that the $j$-th \emph{landmark} of the $i$-th agent particle associates with the MPC set $\mathcal{V}$, the landmark particle weights can be updated by\small
\begin{equation}
\label{landmarkRenew}
    w_{i,j,m}^k=\prod_{\rm{MPC}_n\in \mathcal{V}}\rho_n w_{i,j,m}^{k-1}\frac{1}{\sqrt{2\pi}\sigma_\phi} e^{-\frac{(\hat{\phi}^k-\phi^k)^2}{2\sigma_\phi^2}}\frac{1}{\sqrt{2\pi}\sigma_\tau} e^{-\frac{(\hat{\tau}^k-\tau^k)^2}{2\sigma_\tau^2}},
\end{equation}\normalsize
where $\rm{MPC}_n$ denotes the $n$-th MPC, $\hat{\phi}^k$ and $\hat{\tau}^k$ are the AOA and TOA calculated using the position of the $m$-th landmark particle and the $i$-th agent position, $\sigma_\phi$ and $\sigma_\tau$ denote the corresponding variances. 

$\bullet$ Weight update of agent particles: The weight of the $i$-th agent particle $w_i^k$ is updated by the weights of its landmark particles, which can be written as follows\cite{7487067}:
\begin{equation}
\label{agentRenew}
    w_i^k=w_i^{k-1}\prod_{j=1}^{N_{L.i}}\sum_{m=1}^{N_{lp,i,j}}w_{i,j,m}^k,
\end{equation}
where $N_{L,i}$ is the number of \emph{landmarks} of the $i$-th agent particle.

$\bullet$ Normalize and resample: For all the agent and landmark particles, their weights are normalized. Then, the sets of landmark and agent particles are resampled to discard particles with weights close to $0$. The resample rate of a particle is proportional to its weight.

$\bullet$ Agent and \emph{landmark} update: The pose of the agent in the temporal cycle $\bm{r}_a^k$ can be obtained by the weighted sum of all the agent particles, and the positions of \emph{landmarks} $L^k$ is calculated by weighted sum landmark particles, respectively.

$\bullet$ Map building: The map contains information about positions of reflectors, scatterers and the RIS. The positions of RIS and scatters in the map can be directly obtained from $\mathcal{L}^k$. As for the location of reflectors, the geometry method proposed in \cite{8528841} can be used.

The LME function is summarized in Algorithm \ref{alg:A}.
\begin{algorithm}[t]
\caption{LME function for cycle $k$}
\label{alg:A}
\renewcommand{\algorithmicrequire}{\textbf{Input:}}
\renewcommand{\algorithmicensure}{\textbf{Output:}}
\begin{algorithmic}[1]
\REQUIRE 
         \emph{measurement} $\bm{z}^k$, map $\mathcal{M}^{k-1}$, the pose of agent $\bm{r}_a^{k-1}$, \emph{landmark} set $\mathcal{L}^{k-1}$
\ENSURE 
        Map $\mathcal{M}^{k}$, the pose of agent $\bm{r}_a^{k}$,\emph{landmark} set $\mathcal{L}^{k}$
        \STATE Associate each \emph{measurement} $\bm{z}^k$ with a \emph{landmark} in $\mathcal{L}^{k-1}$.
        \STATE Update the pose $\overline{\bm{r}}_a^{k}$ of agent particles.
        \STATE Calculate the weights $\rho$ for all MPCs.
        \STATE Set the weight $w_{i,j,m}^k$ of all landmark particles using (\ref{landmarkRenew}).
        \STATE Set the weight $w_i^k$ of all the agent particles using (\ref{agentRenew}).
        \STATE Normalize $w_{i,j,m}^k$, $w_i^k$ and resample all particles.
        \STATE Calculate $\bm{r}_a^k$ and $\mathcal{L}^k$ by weighted summation of agent particles and landmark particles.
        \STATE Build map $\mathcal{M}^k$ using updated $\mathcal{L}^k$
\end{algorithmic}
\end{algorithm}

\begin{figure*}[t]
\centering
\subfigure[]{
\begin{minipage}[t]{0.3\linewidth}
\centering
\includegraphics[width=2.2in]{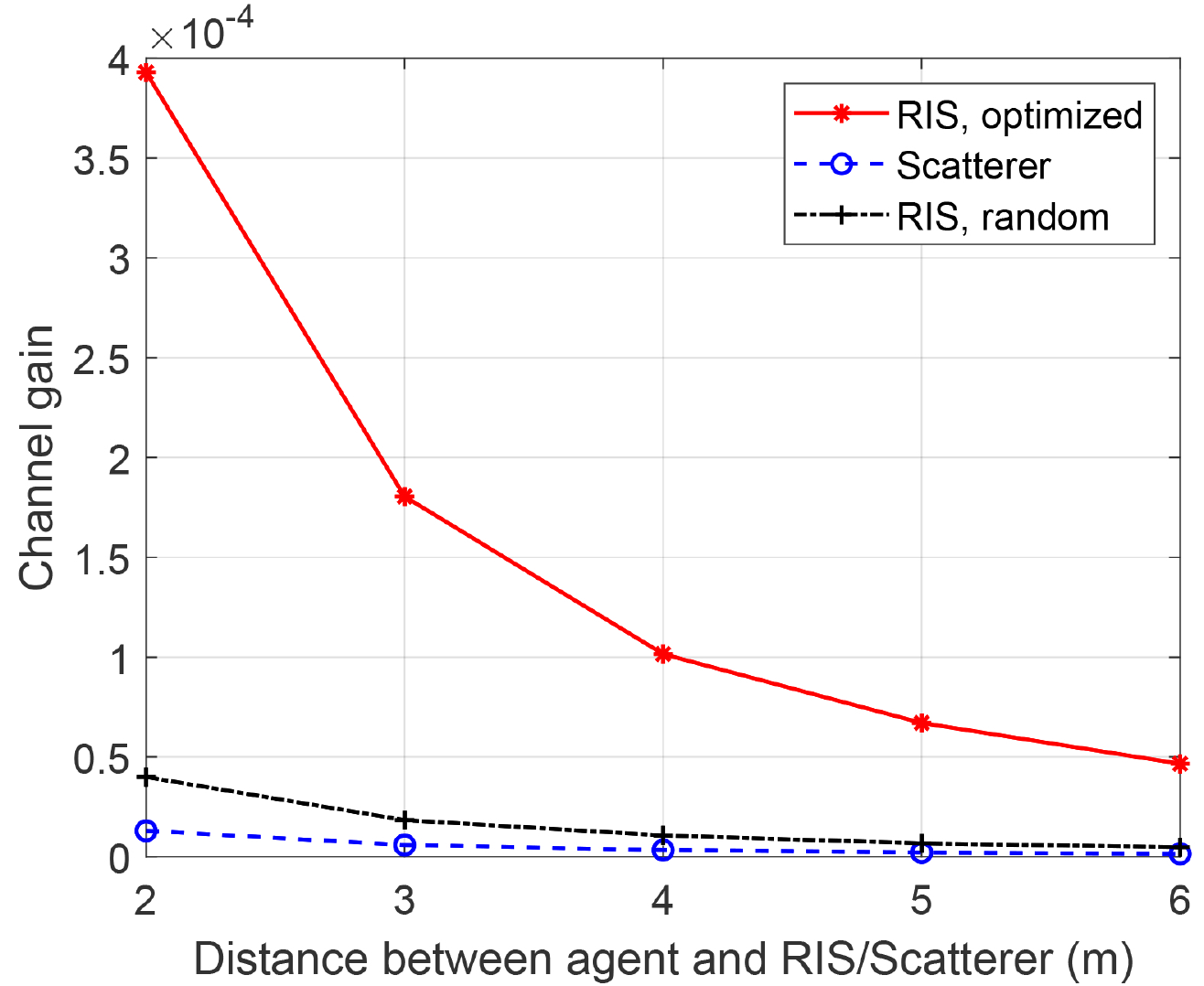}

\end{minipage}%
}%
\subfigure[]{
\begin{minipage}[t]{0.3\linewidth}
\centering
\includegraphics[width=2.2in]{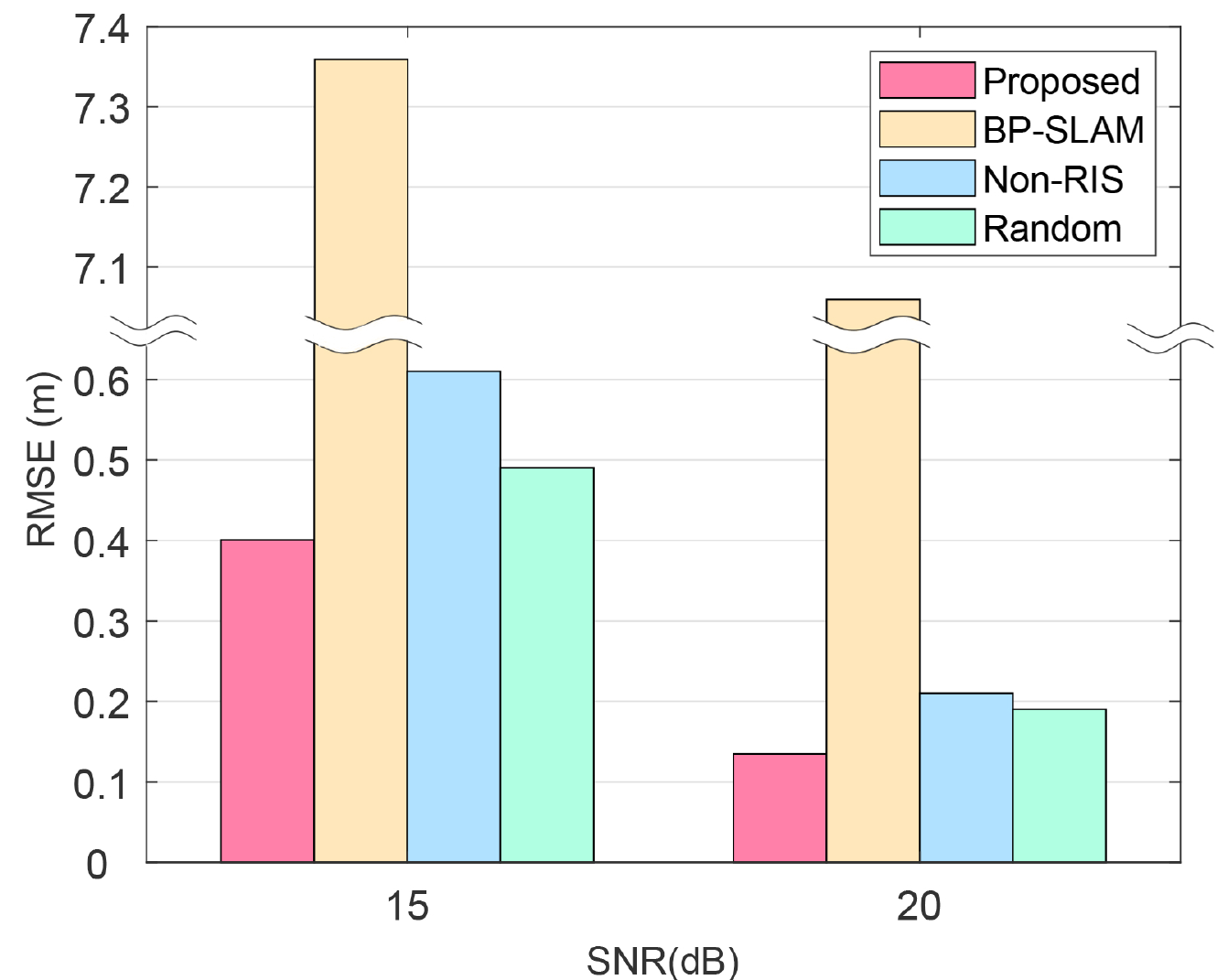}

\end{minipage}%
}%
\subfigure[]{
\begin{minipage}[t]{0.3\linewidth}
\centering
\includegraphics[width=2.2in]{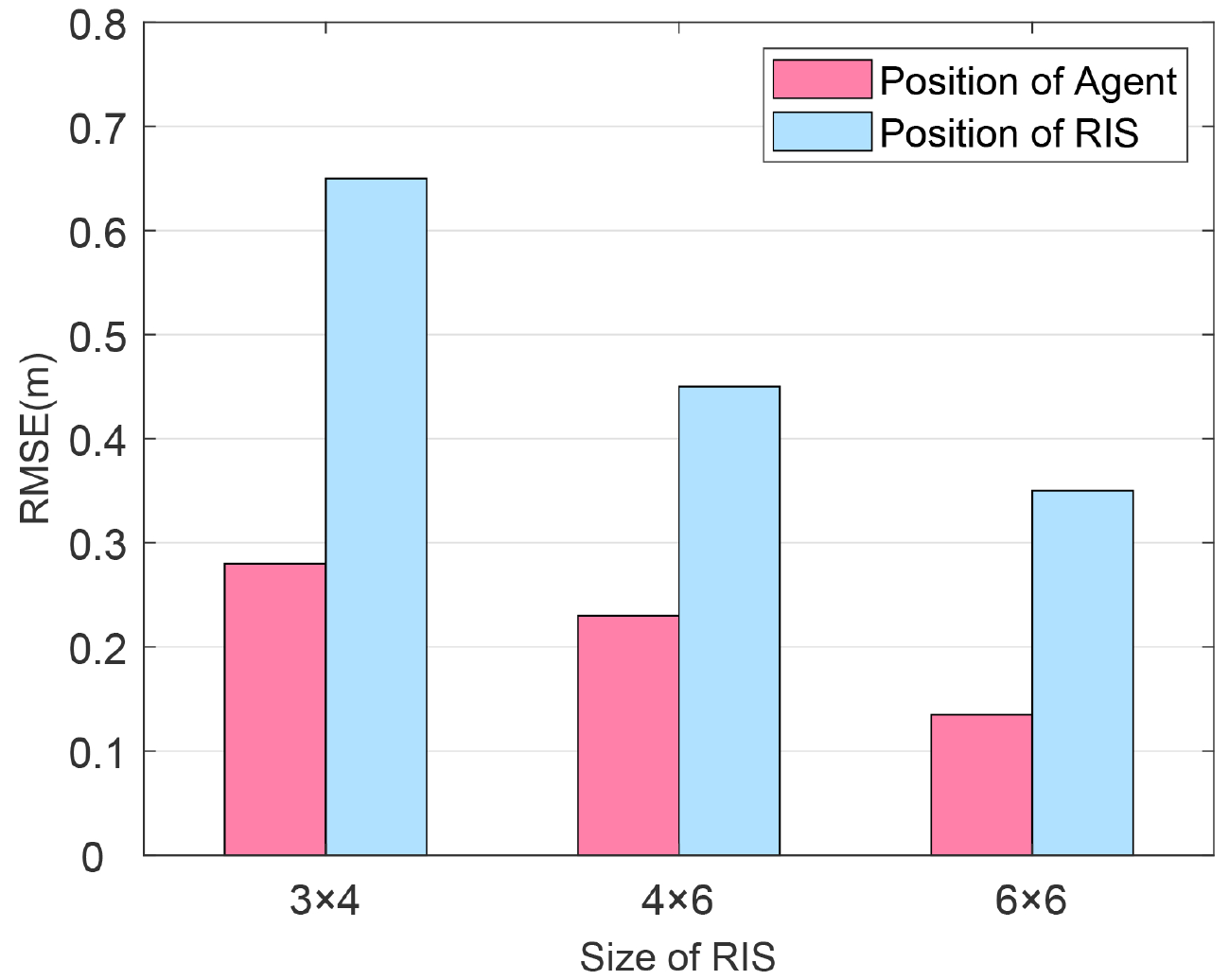}

\end{minipage}
}%
\centering
\caption{(a) Channel gain $\alpha$ versus distance between agent and RIS/Scatterer; (b) Agent RMSE $\varepsilon$ versus SNR; (c) RMSE $\varepsilon$ versus size of RIS}
\end{figure*}
\section{Simulation Results}
\label{sec:sim}
In this section, we present the performance of the proposed SLAM system. The layout of the indoor SLAM scenario is shown in Fig. \ref{fig:scenario}. The size of the room is  6m$\times$6m$\times$3m. The RIS is on the plane $z=3$, and its center is at (3, 3, 3)m. A scatterer is positioned at (2,1,1)m. The reflection coefficient of the reflectors is set to 0.85 and the scatterer is a sphere whose radius is 5cm. The agent starts from (0, 0, 0)m with velocity $\bm{v}=$(0.0707, 0.0707, 0)m/s. The distance between antennas Tx and Rx on the agent is 0.1m, and Tx transmits signal with frequency $f_c$=10~GHz. The RIS is composed of 36 elements, and the element separation is 0.015m. Each element has 4 possible phase shifts with ideal amplitude ratio, i.e. $r$=1. The agent moves for 600 cycles, and each cycle takes 0.1s. The number of particles for agent is 2000, the number of particles for each \emph{landmark} is 6000.

Fig.~4(a) shows the gains of channels $\alpha$ via RIS with optimized phase shifts, random phase shifts, and a scatterer versus the distance between them and the agent. The scatterer is a sphere equal to the size of the RIS, and  is also placed at (3, 3, 3)m. We can observe that the gain of channel via the RIS with optimized phase shifts is significantly larger than those of channels via the RIS with random phase shifts and the scatterer. This indicates that after optimizing the phase shifts of the RIS, the strengths of multipaths via the RIS can be significantly enhanced to improve the SLAM performance.

Fig.~4(b) shows the positioning root mean square error (RMSE) $\varepsilon$ of the agent versus the signal to noise ratio~(SNR). The SNR can be calculated by ${\rm SNR}=\frac{||\bm{Y}(t)||^2}{LU\sigma_n^2}$, where $\bm{Y}(t)$ denotes a vector consists of all samples of signals received by all $L$ antennas, and $U$ denotes the number of samples. Four schemes are considered: RIS with optimized phase shifts, RIS with random phase shifts, non-RIS SLAM, and BP-SLAM\cite{8823946}. We can find that under different SNRs, the agent position RMSEs of all other three schemes are larger comparing to RIS with optimized phase shifts scheme, which means RIS enhanced the performance of the SLAM system.

Fig.~4(c) illustrates the positioning RMSE $\varepsilon$ of the agent and the RIS versus the size of RIS. It can be observed that with the size of RIS goes up, the MMSE of agent decreases, showing that larger RIS can better enhance the performance of SLAM system.

\section{Conclusion}
\label{sec:conclusion}
In this paper, we have designed an RIS-assisted wireless indoor SLAM system. We have proposed channel models of the wireless environment containing RIS. We have introduced and an RIS-aided SLAM protocol to coordinate RIS and the agent. Based on the protocol, we formulated the optimization problem for SLAM. To solve the optimization problem, we have designed a
particle filter based localization and mapping algorithm. It can be concluded from the simulation results that the RIS can significantly enhance the amplitude of channels comparing to scatters. Moreover, the RIS assisted SLAM system can reduce the estimation error of the agent by 0.1m comparing to the non-RIS wireless SLAM system.
\appendices
\section{Derivation of Channel Gain $\alpha_3$}
By applying far-field model, $r_{n,m}^t$ and $r_{n,m,l}^r$ can be expressed as:
\begin{equation}
\label{eq:t}
    r_{n,m}^t=d_{T,RIS}-\bm{z}_{n,m}\cdot\bm{z}_t,
\end{equation}
\begin{equation}
\label{eq:r}
    r_{n,m,l}^r=d_{RIS,R}-\bm{z}_{n,m}\cdot\bm{z}_r,
\end{equation}
where $\bm{z}_{n,m}$ denotes the position of the $(n,m)$-th element, $\bm{z}_t$ and $\bm{z}_r$ denote unit vectors in directions $(\theta_t,\phi_t)$ and $(\theta_r,\phi_r)$.

By taking (\ref{eq:t}) and (\ref{eq:r}) in (\ref{eq:single}) and substituting $(\theta_{n,m}^t,\phi_{n,m}^t)$ and $(\theta_{n,m.l}^r,\phi_{n,m,l}^r)$ with $(\theta_t,\phi_t)$ and $(\theta_r,\phi_r)$, the gain of subchannel~(n,m) can be approximated as
\small
\begin{equation}
\begin{split}
     \alpha^\prime_{n,m}&=\frac{\lambda\sqrt{GG_{RIS}F(\theta_t,\phi_t)F(\theta_r,\phi_r)d_xd_y}}{(4\pi)^{3/2}d_{T,RIS}d_{RIS,R}}\\
    &\times e^{\frac{-j2\pi(d_{T,RIS}+d_{RIS,R}-\bm{z}_{n,m}\cdot\bm{z}_t-\bm{z}_{n,m}\cdot\bm{z}_r)}{\lambda}} Ae^{-j\xi_{n,m}}.
\end{split}
\end{equation}\normalsize
Thus, we can derive (\ref{a3}) by adding the gains of channels via all the elements together.
\section{Proof of Proposition 1}
Since the signals are functions of time delays $\boldsymbol{\tau}^k=[\tau_1^k,...,\tau_{P}^k]^T$ and amplitues, by introducing intermediate variable $\boldsymbol{\chi}^k=[\boldsymbol{\tau}^k,{\rm real}(\boldsymbol{\alpha}^k),{\rm imag}(\boldsymbol{\alpha}^k)]^T$, in this way, FIM $\bm{J}(\boldsymbol{\eta}^k)$can be expressed as
\begin{small}
\begin{equation}
\label{JT}
    \bm{J}(\boldsymbol{\eta}^k)=T \cdot \bm{J}(\boldsymbol{\chi}^k) \cdot T^T,
\end{equation}
\end{small}
where $T$ represents Jacobian matrix, and it is composed of four submatrices:\small
\begin{equation}   \label{T}
T=\frac{\partial\boldsymbol{\chi}^k}{\partial\boldsymbol{\eta}^k}=\left[\begin{matrix}
H_{2\times P}&0_{2\times2P}\\ 0_{2P\times P} & I_{2P \times 2P}
\end{matrix}\right]
\end{equation}\normalsize
where the $l$-th column for $H$ is $\frac{1}{c}[{\rm cos}\nu_l^k,{\rm sin}\nu_l^k]^T$, $\nu_l^k$ is the receiving angle of the $l$-th channel.

Besides, $\bm{J}(\boldsymbol{\chi}^k)$ is the FIM for variance $\boldsymbol{\chi}^k$, following \cite{6363827}, $\bm{J}(\boldsymbol{\chi}^k)$ can be written as\footnotesize
\begin{equation}
\label{J}
   \bm{J}(\boldsymbol{\chi}^k)=\left[\begin{matrix}\Lambda_A & \Lambda_B^R & \Lambda_B^I\\(\Lambda_B^R)^T & \Lambda_C & 0 \\(\Lambda_B^I)^T & 0 & \Lambda_C\end{matrix}\right].
\end{equation}\normalsize
The elements of $\bm{J}(\boldsymbol{\chi}^k)$ is given by\footnotesize
\begin{align}
    [\Lambda_A]_{l,l^\prime}=&\frac{2}{N_0}\mathcal{R}\{\alpha_l^k(\alpha_{l^\prime}^k)^{*}\}\frac{\partial^2R_s(\tau^k_l-\tau^k_{l^\prime})}{\partial\tau^k_l\partial\tau^k_{l^\prime}},\\
    [\Lambda_B^R]_{l,l^\prime}=&\frac{2}{N_0}(\alpha_l^k)^R\frac{\partial^2R_s(\tau^k_l-\tau^k_{l^\prime})}{\partial\tau^k_l},\\
    [\Lambda_B^I]_{l,l^\prime}=&\frac{2}{N_0}(\alpha_l^k)^I\frac{\partial^2R_s(\tau^k_l-\tau^k_{l^\prime})}{\partial\tau^k_l},\\
    [\Lambda_C]_{l,l^\prime}=&\frac{2}{N_0}R_s(\tau_l^k-\tau_{l^\prime}^k),
\end{align}\normalsize
where $R_s(\tau)= \int_{-\infty}^{\infty}s(t)s(t-\tau)dt$. By taking (\ref{J}) and (\ref{T}) into (\ref{JT}), we can get (\ref{derived}).


%





\ifCLASSOPTIONcaptionsoff
  \newpage
\fi

\end{document}